\documentclass[proceedings]{JHEP}
\conference{Heavy Flavours 8, Southampton, UK, 1999}
\usepackage{epsfig}
 
\title{Charm counting and $B$ semileptonic branching fraction }
\author{Hitoshi Yamamoto \\ University of Hawaii , 2505 Correa Rd., Honolulu HI 96822
USA
\\
 E-mail: hitoshi@phys.hawaii.edu}
\abstract{ We review recent measurements 
as well as phenomenological background ofthe  semileptonic branching
fraction of $b$ hadrons and number of charm produced 
per decay of $b$ hadrons.
 }

\def\Bsl{B_{\rm s.l.}}
\def\Gsl{\Gamma_{\rm s.l.}}
\def\Ghad{\Gamma_{\rm had}}
\def\rhad{r_{\rm had}}
\def\rrare{r_{\rm rare}}
\def\rud{r_{\bar ud}}
\def\rcs{r_{\bar cs}}

\def\ovbr{\overbrace}
\def\tbst#1#2{\hbox{\vrule height #1 pt depth #2 pt width 0 pt}}

\def\arrvsp{\vspace{5 pt}}
\def\unin#1{\tbst{10}{20}_{\makebox[0pt]{\hss$\ovbr{#1}$\hss}}}
\def\lrlap#1{\hbox to 0pt{\hss#1\hss}}

\def\bra{\langle}
\def\ket{\rangle}

\def\beq{\begin{equation}}
\def\eeq{\end{equation}}
\def\beqa{\begin{eqnarray}}
\def\eeqa{\end{eqnarray}}

\def\Leff{{\cal L}_{\rm eff}}

\begin{document}

\section{Introduction}

The puzzle of inclusive non-leptonic $B$ decay was first pointed out
in around 1994~\cite{BBSV,FWD} when the theoretical prediction
was considered to be difficult to accommodate 12\%\ or less for the
semileptonic branching ratio $\Bsl$ of $B$ meson while 
the measurement published
by the CLEO collaboration was~\cite{CLEOBsl96}
\beq
  \Bsl = 10.49\pm0.46{\rm\%}\quad(\Upsilon(4S)).
  \label{eq:CLEOBsl}
\eeq 
The discrepancy was
particularly stark when viewed in the two-dimensional plane of
$n_c$ vs $\Bsl$, where $n_c$ is the average number of charm or
anticharm created per $B$ decay~\cite{BBSV,FWD,DISY}. 
The experimental value of $n_c$ was~\cite{CLEOnc}
\beq
   n_c = 1.10 \pm 0.05\,.
  \label{eq:CLEOnc}
\eeq
while the naive expectation was around 1.16.
This was because
when the uncertainties of the theory, in particular the charm quark mass,
were tweaked to reduce $\Bsl$, it increased the inclusive non-leptonic decay
rates which resulted in too large a value for $n_c$ compared to measurement.
We have now new measurements related to this subjects which we will review
below together with phenomenological background.

\section{Definitions}

By $\Bsl$, we mean the average branching fraction for direct electron
production. The average is taken over the hadrons containing $b$ quark
produced in a given environment. 
It is usually assumed that the electronic branching fraction is the same
as the muonic branching fraction.
In reality $\Bsl$ is average over weakly-decaying hadrons containing one
$b$ quark.
In the case of experiments running on
the $\Upsilon(4S)$ resonance, the average is over $B^+$ and $B^0$
and their charge conjugate hadrons~\footnote{Charged conjugate hadrons
and decay channels are implicit in the following.}
produced nearly equal amount, and for the experiments running on\
$Z^0$, the average is taken over $B^+$, $B^0$, $B_s^0$, and baryons
containing one $b$ quark which we denote as $N_b$. 
$N_b$ is in turn the mixture of $\Lambda_b (udb)$, 
$\Sigma_b (usb)$, $\Xi_b (dsb)$, and
$\Omega_b (ssb)$.
The relative fractions
are roughly $B^+$:$B^0$:$B_s^0$:$N_b$ = 4:4:1:1; or more precisely~\cite{PDG}
\[
  \begin{array}{r@{\qquad\arrvsp}l}
    B^+ & 39.7^{+1.8}_{-2.2} {\rm\%} \\
    B^0 & 39.7^{+1.8}_{-2.2} {\rm\%} \\
    B_s^0 & 10.5^{+1.8}_{-1.7} {\rm\%} \\
    N_b & 10.1^{+3.9}_{-3.1} {\rm\%} \\
  \end{array}
\]

The charm count $n_c$ is the number of weakly-decaying charm or
anticharm hadrons
produced in the decay of one $b$-hadron, and it is again the average
over the $b$-hadrons produced in the given environment.
one usually counts the total number of $D^+$, $D^0$, $D_s^+$,
$\Lambda_c$, $\Xi_c$.
One exception is chamonium which is counted as two charms if it decayed
by $c\bar c$ annihilation. Namely, a $J/\Psi$ meson produced in $b$-hadron decays
are counted as two, while $\Psi''$ which decays predominantly to $D\bar D$
is counted when $D$ mesons are counted. 

\section{Theoretical tools}

The basis of the predictions for $\Bsl$ and $n_c$ is the assumption is
quark hadron duality which essentially states that the sum of rates to
hadronic final states with a given flavor quantum number is the same as
the sum of the rates at quark level to the same quantum numbers.
There are two versions of this assumption: one is the global duality which
applies to the case where the relevant rates are averaged over some
range of c.m. energy. An example is the semileptonic decay of a $b$ hadron
where the hadronic system in the final state
can have  different c.m. energies.
Namely, in $b\to c \ell\nu$, the $c$ quark in the final state and the
spectator quark that used to be in the parent $b$-hadron will form the
hadronic system of the final state, and the c.m. energy of the system has
a distribution over some range.
The inclusive semileptonic rate is then estimated by
taking the integration over it. 

Another is the local duality where the
rates of real hadronic final states are the same as the quark-level
rates even if the c.m. energy of the hadronic system is fixed.
This assumption is required to calculate the inclusive non-leptonic
rates where all particles in the final states are hadrons.
This is considered to be a stronger assumption than the
global duality; since there are three quarks plus the spectator quark in the
final state, however, one hopes that sufficient averaging over is
involved to make the quark-level calculation reliable.

Then, the quark-level estimation is usually performed in the
the heavy-quark expansion~\cite{HQEX} 
and the perturbative QCD in the framework of the operator-product
expansion~\cite{OPE}. 
The specific application  starts with
the optical theorem for the partial decay rate
\beqa
  \Gamma(B \to f) &=&  \\
  &&\hspace{-0.5in}
   {1\over m_B}{\rm Im}\bra B|i\int d^4x
   T(\Leff(x)\unin{\sum_f|f\ket\bra f|} \Leff(0))|B\ket\,.
   \nonumber 
\eeqa
After expansion in terms of $1/m_b$, one obtains
\beqa
  \Gamma(B\to f) &=& \Gamma_0 \Big[
    a \Big(1+{\lambda_1\over 2m_b^2}
     + {3\lambda_2\over 2m_b^2}\Big) \nonumber \\
   &&  b {\lambda_2\over m_b^2} + {\cal O}({1\over m^3_b})\Big]\,,
\eeqa
where $\Gamma_0 = G_F^2m_b^5/192\pi^3$, and
the coefficients $a,b$ contains the effect of phase space,
the Wilson coefficients estimated by perturbative QCD, and the
CKM factors. The parameters $\lambda_1$ and $\lambda_2$ incorporates
non-perturbative effects. The parameter $\lambda_1$ corresponds to
$-\bra \vec p_b^2 \ket$ where $\vec p_b$ is the fermi-motion momentum
of the $b$ quark inside the hadron, and one can identify the factor
$1+\lambda_1/2m_b^2$ as the time delation factor for the decaying $b$
quark. There is a large uncertainty associated with the definition
as well as the estimation of $\lambda_1$ varying from 0 to $\sim -0.7$
GeV$^2$. However, the effect on the non-leptonic decay rate is less than
2 \%. The chromo-magnetic effect is contained in the parameter
$\lambda_2$ which can be reliably estimated from the splitting of
$B^*$ and $B$ mesons:
\beq
   \lambda_2 = {m_B^{*2} - m_B^2\over 4} \sim 0.12 {\rm GeV}^2\,.
\eeq
Overall perturbative effect included in $\lambda_{1,2}$ is
less than a few \%. There are, however, non-perturbative effects that
are not included in $\lambda_{1,2}$ such as possible large final-state
interactions in $b\to c\bar cs$ mode where the particles in the final state
are moving quite slowly.

The semileptonic branching fraction is then estimated as
\beq
   \Bsl = {\Gsl\over 2.22\Gsl + \Ghad}
     = {1\over 2.22 + \rhad}\,,
\eeq
where the factor 2.22 accounts for the total rate of
$\ell\nu X$ $(\ell = e,\mu,\tau)$ in unit of $\Gsl$
(1 for $e$, 1 for $\mu$, and 0.22 for $\tau$), and 
$\rhad$ is the total non-leptonic rate again in unit of $\Gsl$:
\beq
   \rhad = \rud + \rcs + \rrare\,.
\eeq
Here, $\rud$ ($\rcs$) is the rate for the decays caused by
$b\to c\bar ud'$ ($b\to c\bar cs'$) transitions 
where $d'$ ($s'$) is the appropriate Cabibbo mixture of
$d$ and $s$ quarks, and $\rrare$ includes all charmless hadronic 
decays including penguins and $b\to u$ transitions. 
The value of $\rrare$ is estimated in the standard model to be~\cite{BDY,rrare}
\beq
    \rrare = 0.25\pm0.10\,.
  \label{eq:rrare}
\eeq
The decays caused by $b\to u\bar c s'$ transition is not included in
the above classifications, but it is quite small at around 0.4\%\
branching fraction.

A complete next-to-leading calculation of $\rud$ has been 
performed~\cite{rudNLO}
which gives
\beq
      \rud = 4.0\pm0.4\,,
  \label{eq:rudNLO}
\eeq
where naively we expect $3$ for the color factor. The uncertainties
involved are the renormalization scale $\mu$, quarks masses $m_c$ and
$m_b$, and the assumption of quark hadron duality. The quark masses are
constrained by the heavy-quark expansion relation
\beqa
   m_b - m_c &=& m_B - m_D  \nonumber \\
   &&\hspace{-0.6in}+\; {\lambda_1 + 3\lambda_2\over 2}
   \Big( {1\over m_b} - {1\over m_c} \Big) + {\cal O}({1\over m_b^3})
      \nonumber \\
   &&\hspace{-0.6in} \sim 3.40\pm0.06 {\rm GeV}\,.
\eeqa
Thus, in the following, it is understood that changing $m_b$ will
change $m_c$ accordingly. The value of $\rud$ as a function of $\mu/m_b$
is shown in the figure~\ref{fg:rud} for $m_b = 4.5$ GeV. 
\EPSFIGURE[h]{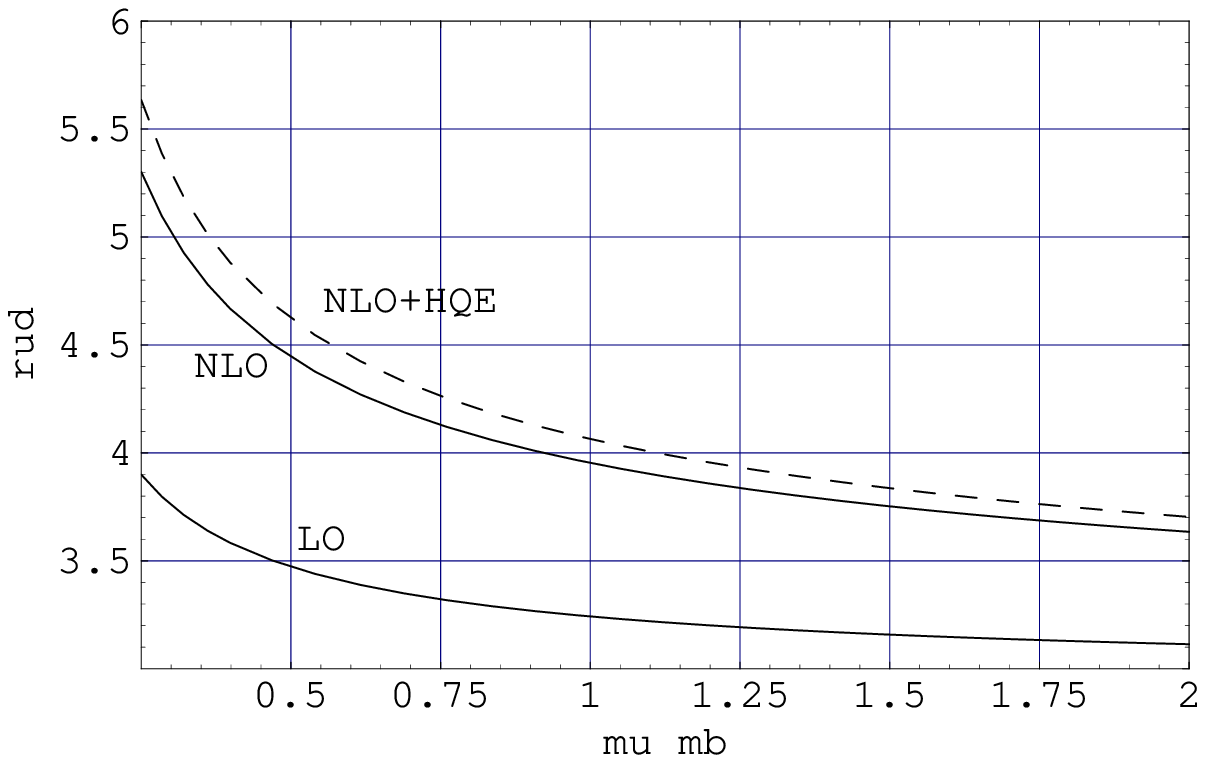,width=7cm}{The estimate of $\rud$
 for LO, NLO, and NLO + HQE estimations.}
One notes that there is a substantial difference between the leading
order (LO) estimation and the next-to-leading (NLO) estimation, and that
the scale dependence of the NLO is not much better than that of LO.
Both features are not encouraging with respect to the reliability of
the final number. The correction due to the nonperturbative
effects that are within the framework of the heavy-quark expansion (HQE)
are small as mentioned earlier.

The uncertainty associated with the estimation of $\rcs$ is generally
considered to be larger than that for $\rud$. It is more sensitive to the
quark masses and it was realized that there is a large QCD correction
with finite charm quark mass
due to a hard gluon emission from one the charm quarks in the final 
state~\cite{ncbagan,voloshin}. The correction enhanced the rate by about 30\%.
Also, the slow velocity of charm quarks makes it susceptible to
final-state interactions~\cite{DISY}.

Since the $b\to c\bar us'$ mode has one charm, the $b\to c\bar s'$ has
two charms, and the 'rare' mode has no charm,
the number of charm per $B$ decay $n_c$ is given by
\beqa
  n_c &=& 1 + Br_{\bar cs} - Br_{\rm rare} \nonumber \\
   &=& 1 + {\rcs - \rrare \over 2.22 + \rhad}\,.
\eeqa
The theoretical values for $\Bsl$ and $n_c$ are~\cite{NS}
\beq
  \begin{array}{c@{\arrvsp}cc}
    \mu/m_b               & 1             &      1/2        \cr
     \Bsl({\rm\%}) \quad  &  12.0\pm1.0   &    10.9\pm1.0   \cr
      n_c                 &  1.20\pm0.06  &   1.21\pm0.06   \cr
  \end{array}
\eeq
The quark masses used is $m_c/m_b = 0.29$ or equivalently $m_b \sim 4.8$ GeV.
It is seen that when $\mu$ is decreased $\Bsl$ goes down while $n_c$ is 
relatively stable; on the other hand, as $m_c/m_b$ is reduced 
$b\to c\bar c s'$ rate increases with respect to $\Gsl$ and
as the result $\Bsl$ goes down and $n_c$ goes up.
If we would like to do away with the uncertainty in the $b\to c\bar c s'$ mode,
one could eliminate $\rcs$ from the expression of $\Bsl$ and $n_c$ and
obtain
\beq
   n_c = 2 - (2.22 + \rud + 2\rrare) \Bsl\,,
\eeq
which is a linear relation between $n_c$ and $\Bsl$ and the dominant error is
that in $\rud$ if we are to trust the standard estimation of $\rrare$. Using the
values for $\Bsl$ (\ref{eq:CLEOBsl}), $\rud$ (\ref{eq:rudNLO}),
and $\rrare$ (\ref{eq:rrare}), $n_c$ becomes $1.30 \pm 0.06$ which is
to be compared with $1.10\pm0.05$ (\ref{eq:CLEOnc}). This 2.5 $\sigma$
discrepancy largely prompted proposals of new physics 
that boosts $\rrare$~\cite{BBSV,FWD,newphys}.

Sizes of corrections that affect $\Bsl$ and $n_c$ are shown in
the table~\ref{tb:bslnccorr}.
\TABULAR[h]{cccc}{
        & naive & NLO($m_c = 0$) & NLO($m_c \not= 0$)\arrvsp \\
      $r_{\ell\nu}$ &  2.22 &         2.22 &      2.22 \\
      $\rud$        &  3.0  &         4.0  &      4.0  \\
      $\rcs$        &  1.2  &         1.6  &      2.1  \arrvsp\\
      $\Bsl$        &  0.16 &        0.13  &      0.12 \\
      $n_c $        &  1.16 &        1.17  &      1.21 \\
      $Br_{\rm cs}$ &  0.18 &        0.20  &      0.25 
                 }
     {Effects of various corrections affecting $\Bsl$ and $n_c$.
     \label{tb:bslnccorr}}
After all the corrections, the theoretical value of $\Bsl$ has come down
and more or less consistent with the measurement. However, it is 
accomplished by boosting $\rcs$ and it increased the estimation of
$n_c$. 

\section{Measurements of $\Bsl$}

On $\Upsilon(4S)$, the state-of-the-art is the correlated di-lepton
method~\cite{ARGUS2lep}
where one $B$ is tagged by a high-momentum lepton and a lepton
is searched for on `the other side.¹ This allows one to separate the direct
$b\to \ell^-$ from the cascade $b\to c\to \ell^+$ using the charge correlation
of the tagged-side lepton and the signal-side lepton. 
This reduces the model dependence due to the subtraction of the cascade
component.
The effect due to
the $B^0-\bar B^0$ mixing can be unfolded in each momentum bin by
solving
\beq
   \begin{array}{rl}
       {dN_{+-}\over dp} &= N_\ell \epsilon
       \left[ {dB_b\over dp} (1 - \chi) + {dB_c\over dp} \chi \right] 
         \arrvsp \\
       {dN_{\pm\pm}\over dp} &= N_\ell \epsilon
       \left[ {dB_b\over dp} \chi + {dB_c\over dp} (1 - \chi) \right] 
   \end{array}
\eeq
where $\chi = 0.080\pm 0.012$ is the mixing parameter, $N_{+-}$ ($N_{++}$)
is the observed number of opposite-sign (same-sign) dileptons,
$N_\ell$ is the total number of tagging leptons, $\epsilon$ is the
lepton detection efficiency, and $dB_b/dp$ ($dB_c/dp$) is the
direct (cascade) lepton spectrum from $B$.
The measurement (\ref{eq:CLEOBsl}) was made with this technique.
The spectra thus obtained is shown in the figure~\ref{fg:CLEOBsl}.
\EPSFIGURE[h]{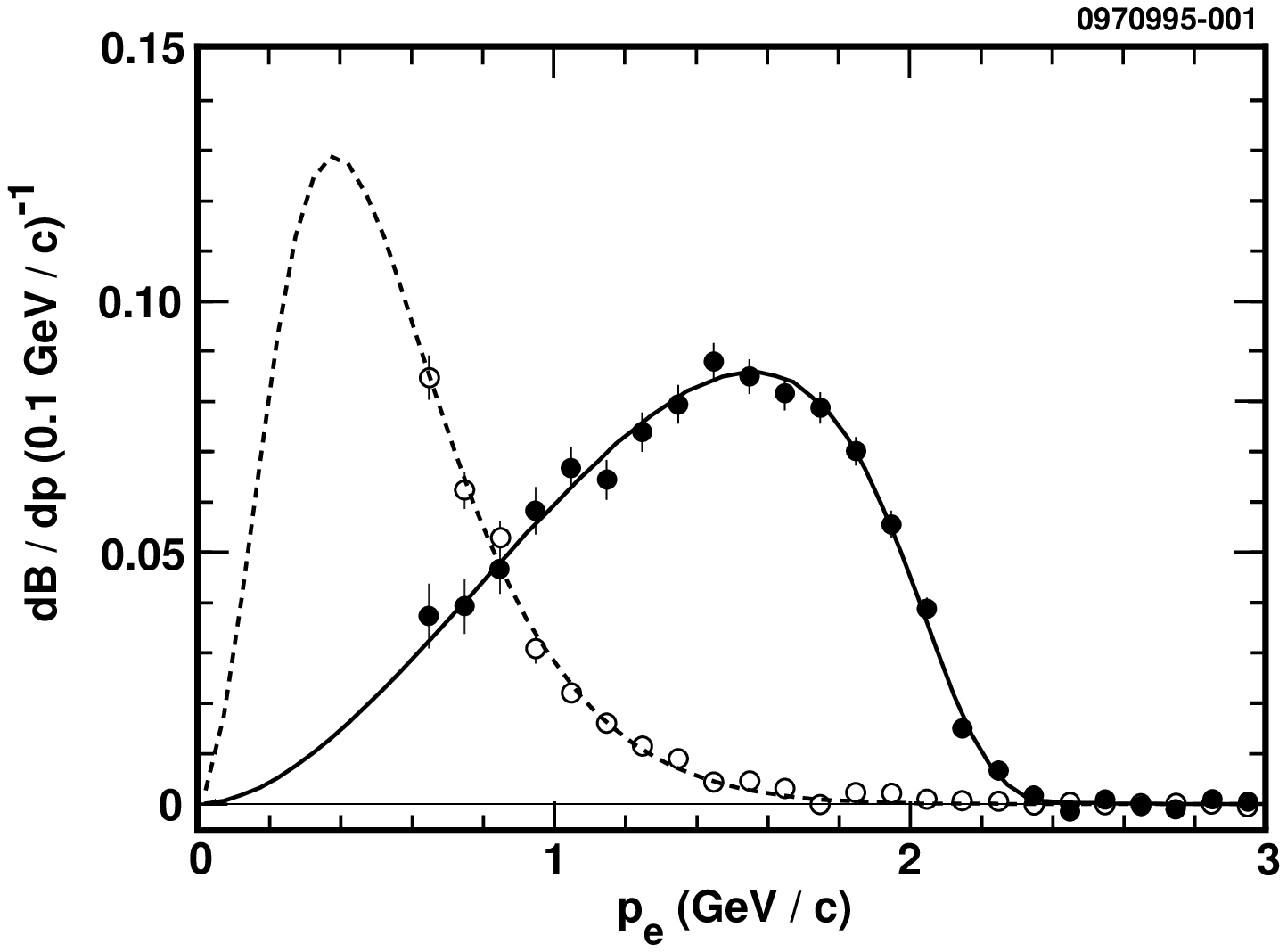,width=7cm}{Unfolded
      spectra for direct (black circles) and cascade leptons 
      (open circles) by the CLEO collaboration.
      \label{fg:CLEOBsl}}
In principle, the leptons from the wrong-sign cascade
$b\to \bar c \to \ell^-$, which can occur through $b\to c\bar cs$
where the $c\bar c$ pair fragment to $D\bar D_{(s)} X$, can contaminate the
direct lepton sample. The momentum of such leptons, however, is
low enough that the effect was found to be negligible.

The table~\ref{tb:LEPBsl} shows recent dedicated
measurements of $\Bsl$ on the $Z^0$ peak. 
Values of $\Bsl$ obtained in global fits to electro-weak parameters
are not included in this list.
\TABULAR[h]{cccc}{
      Experiment  & $\Bsl$(\%) \arrvsp \\
      ALEPH 95~\cite{ALEPHBsl}  &  $11.01\pm0.10\pm0.30$ \\
      L3 96~\cite{L3Bsl}        &  $10.68\pm0.11\pm0.46$  \\
      OPAL 99~\cite{OPALBsl}    &  $10.83\pm0.10\pm0.20^{+0.20}_{-0.13}$ 
       \arrvsp \\
      DELPHI 99~\cite{DELPHIBsl}&  $10.65\pm0.07\pm0.25^{+0.28}_{-0.12}$  \\
                 }
     {Recent measurements of $\Bsl$ at LEP. The model dependence is separated
      out as the last error for OPAL and DELPHI.
     \label{tb:LEPBsl}}
The Aleph analysis used a charge correlation method similar
to the CLEO measurment above while single lepton sample was also used.
\EPSFIGURE[h]{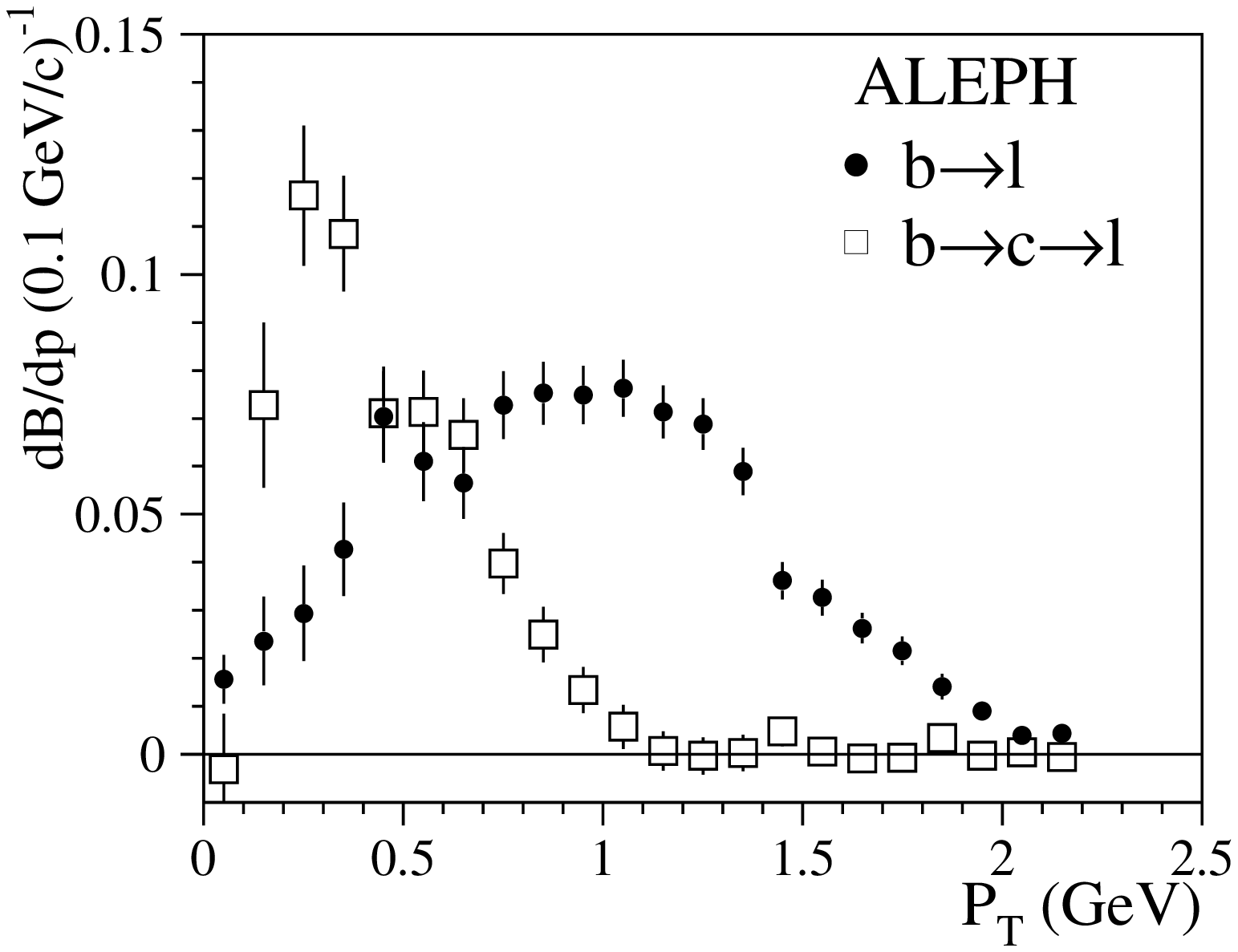,width=6.7cm}{The
      $p_T$ pectra for direct and cascade leptons 
      by the ALEPH collaboration. \label{fg:ALEPHBsl}}
The lepton spectrum obtained by ALEPH is shown in the figure~\ref{fg:ALEPHBsl}.
The spectrum is shown as functions of $p_T$ with respect to jet axis,
and this makes the analysis sensitive to the contamination from
the wrong-sign cascade $b\to \bar c \to \ell^-$.

The measurement by the L3 collaboration~\cite{L3Bsl} combines separate measurements
of $b\to e\nu X$, $\mu\nu X$, and $\nu X$. The neutrino mode is detected
by large missing energy. The final number for $\Bsl$ is then obtained by
a fit to the three samples. Charge correlations are not used.

The $\Bsl$ measurement reported by OPAL~\cite{OPALBsl} 
begins by enhancing $Z^0\to b\bar b$
events by a lifetime flavor tagging technique based on neural net.
The $b$-hemisphere sample is thus selected with a purity of 92\%\ and
an efficiency of 30\%. Lepton is searched in the
jet opposite to the tagged side. Two neural net variables $NN_{b\ell}$
and $NN_{bc\ell}$ are formed using $(p,p_T)$ of the lepton,
energy of the lepton-side jet, charge correlation of the lepton and the
jet containing the lepton and that of the lepton and the most energetic jet
on the tagging side, and impact parameter of the lepton. Also, the energy
of the subjet containing the lepton is also used to enhance the sensitivity
to $b\to c\to \ell^+$, and the scalar sum of $p_T$ of the lepton jet is used
to enhance the sensitivity against light quark jets. Then, a maximum likelihood
fit is performed on $NN_{b\ell}$
and $NN_{bc\ell}$ to extract number of direct leptons and cascade leptons.
\EPSFIGURE[h]{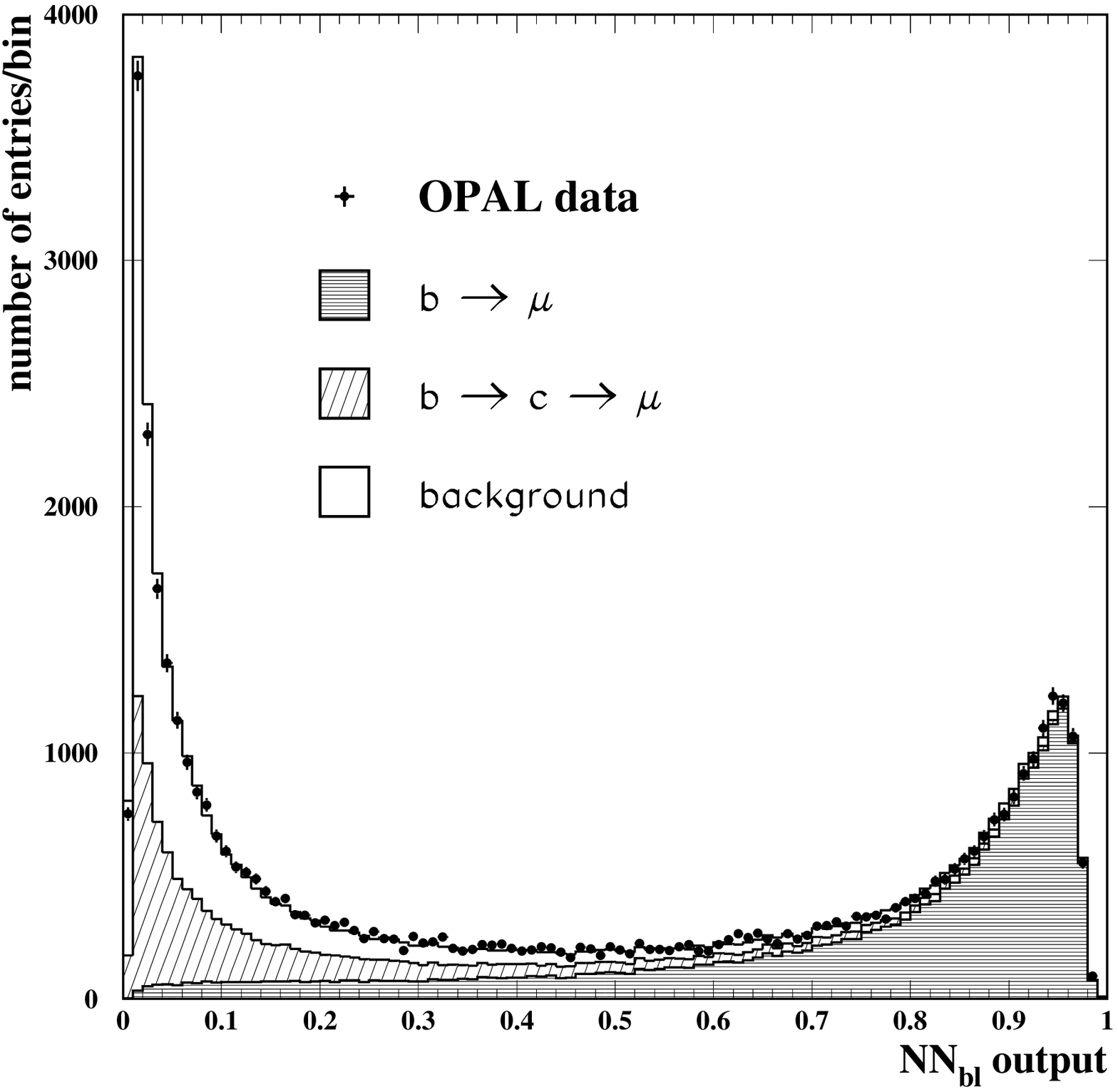,width=6.7cm}{The
      result of maximum likelihood fit to the neural
      net variable $NN_{bl}$
      by the OPAL collaboration. \label{fg:OPALBsl}}
The result of the fit is shown in the figure~\ref{fg:OPALBsl}.
There are substantial difference in distribution shape for
the three components which allows the separation among them.
In addition to the direct $b\to \ell$ branching fraction,
the fit also gives the cascade lepton branching ratio:
\beq
   Br(b\to c\to \ell^+) = 8.40\pm 0.16\pm 0.21^{+0.33}_{-0.29}{\rm\%}\,.
\eeq
Both for $\Bsl$ in the table as well as in the above, the last error
represents model dependencies since the fit does require the
lepton spectra for $b\to \ell$ and $b\to c\to\ell$ as well as
fragmentation functions. The inclusion of the charge correlations in the
fit, however, reduces the systematic error due to model dependences.

The DELPHI collaboration has combined four independent analyses:
\begin{enumerate}
\item The traditional single lepton counting on the opposite side to
   a hemisphere $b$-tagged by vertexing and lepton without
   charge correlation, plus di-lepton sample with charge correlation.
   Fitting the $(p,p_T)$ distribution of the lepton(s), 
   $\Bsl$ is extracted as
   \beq
       \Bsl = 10.66\pm0.11\pm0.24^{+0.25}_{0.12}{\rm\%}\,.
   \eeq
\item The $b$-hemisphere is tagged by vertexing and lepton on
  the opposite side, then two parameters are formed:
  \beq
    \begin{array}{rcl}
    \lambda_Q &=& (\hbox{tag-side jet charge}) \\
       &&\times(\hbox{lepton charge}) \arrvsp\\
    k^* &=& p_{\rm lepton} \hbox{in the c.m. of $b$-hadron}
    \end{array} \nonumber
  \eeq
  The $b$ vertex information
  is used to move to the c.m. of the $b$-hadron in order
  to obtain the absolute lepton momentum there since it has better
  separation power than $p_T$ for the cascade leptons. Then, 
  a fit to the $(\lambda_Q, k^*)$ distribution gives
  \beq
    \Bsl = 10.74\pm0.13\pm0.41^{+0.46}_{0.30}{\rm\%}\,.
  \eeq
\item The analysis uses all hadronic events and employ a
  multi-variate method to separate flavors in which
  $(p,p^{\rm in}_T,p_T^{\rm out})$ of leptons are reconstructed for
  each flavor where $p^{\rm in}_T,p_T^{\rm out}$ are the $p_T$ with and
  without the lepton in calculating the jet axis. The result is
  \beq
    \Bsl = 10.64\pm0.11\pm0.25^{+0.37}_{-0.44}{\rm\%}\,.
  \eeq
\item In this analysis, the $b$ vertex is identified and then
   the charge of $b$ is determined by a neural net using
   jet charges and charged kaons etc. Then, the lepton momentum
   in the $b$-hadron c.m. system is fit separately for two
   relative charges of leptons giving
   \beq
    \Bsl = 10.81\pm0.12\pm0.26^{+0.35}_{-0.52}{\rm\%}\,.
   \eeq
\end{enumerate}
The value of $\Bsl$ in the table is the combined result of the
four measurements above where correlations among the measurements
are taken into account. Also obtained are
\beqa
   Br(b\to c\to\ell^+) &=& 7.88\pm0.13\pm0.27^{+0.38}_{-0.32} {\rm\%}
     \nonumber \\
   Br(b\to c\to\ell^-) &=& 1.71\pm0.13\pm0.36^{+0.25}_{-0.19} {\rm\%}
     \nonumber
\eeqa

Averaging the $\Bsl$ values of the four LEP experiments, the $\Bsl$
on $Z^0$ is
\beq
   \Bsl = 10.79 \pm 0.17 {\rm\%}\quad(Z^0)\,.
  \label{eq:LEPBsl}
\eeq

\section{Charm counting}

There have been two types of charm counting reported so far.
One way is to use the vertex information without explicitly
reconstructing exclusive charm decays. The more traditional
method is to exclusively reconstruct each charm hadrons.
Recent results of the latter method are the
CLEO result (\ref{eq:CLEOnc}) and
\beq
  n_c =
  \left\{
  \begin{array}{c}
    1.23 \pm 0.036\pm 0.028\pm 0.053 \\
     \hbox{(ALEPH 96)} \\
    1.166\pm 0.031\pm 0.059\pm 0.054 \\ 
     \hbox{(DELPHI 99)} \\
  \end{array}
        \right.\,,
  \label{eq:ncLEPexcl}
\eeq
measured on $Z^0$. 
The last error reflects the uncertainty in the branching
fractions of charmed hadron decays.
In the ALEPH analysis, the vertex $b$ tag was
used to enhance $Z\to b\bar b$ events, and then the exclusive
charm decays were counted. For DELPHI, the two sources of charm,
$b\to c$ and $Z\to c\bar c$, were separated by the energy distribution
of the charmed hadron and the vertex information, then the charm count
was extracted using the measured value of $R_b$ which gives
the total number of $Z\to b\bar b$ events.
The breakdown the charm count is given in the table~\ref{tb:ncexcl}
together with results from OPAL which did not give the total
count.
\TABULAR[t]{ccccc@{\arrvsp}}{
  (\%)  & CLEO 96~\cite{CLEOnc}
        & ALEPH 96~\cite{ALEPHnc}
        & OPAL 96~\cite{OPALnc}
        & DELPHI 99~\cite{DELPHInc}   \arrvsp \\
 $D^0$  & 63.6$\pm$3.0 & 60.5$\pm$3.6 & 53.5$\pm$4.1 & 60.05$\pm$4.29 \\
 $D^+$  & 23.5$\pm$2.7 & 23.4$\pm$1.6 & 18.8$\pm$2.0 & 23.01$\pm$2.13 \\
 $D_s^+$& 11.8$\pm$1.7 & 18.3$\pm$5.0 & 20.8$\pm$3.0 & 16.65$\pm$4.50 \\
 $\Lambda_c$
        &  3.9$\pm$2.0 & 11.0$\pm$2.1 & 12.5$\pm$2.6 &  8.90$\pm$3.00 \\
 $\Xi_c^{0.+}$
        &  2.0$\pm$1.0 &  6.3$\pm$2.1 &     $-$      &  4.00$\pm$1.60 \\
 $(c\bar c)\times 2$
        &  5.4$\pm$0.7 &  3.4$\pm$2.4 &     $-$      &  4.00$\pm$1.29 \\
                            }
     {Breakdown of charm counting by exclusive reconstruction of
      charm hadrons. Charmonia are indicated as $(c\bar c)$.
      The CLEO
      result is the average over $B^0$ and $B^+$ while the LEP results
      are the average over $B^0$, $B^+$, $B_s$, and bottom baryons.
     \label{tb:ncexcl}}

Some comments are in order. First,
the CLEO result is the average over $B^0$ and $B^+$ while the LEP results
are the average over $B^0$, $B^+$, $B_s$, and $N_b$ (bottom baryons). Thus
one expects that the LEP results should have larger values for
$D_s^+$ and charmed baryons; this is verified by the measurements.

Charmonia are indicated by $(c\bar c)$ in the table, and it refers
to the $c\bar c$ annihilating portion of 
$J\Psi$, $\Psi'$, $\chi_{0,1,2}$, $\eta_c$, and $h_c$.
Of which $J\Psi$, $\Psi'$, and $\chi_1$ are actually detected, and
theoretical prediction is used for $\eta_c$.
If factorization works, $\chi_1$ and $h_c$ are not expected to be
produced by $V-A$ interaction. For $\chi_{c2}$, CLEO has used
its own measurement $Br(B\to \chi_{c2}) = 0.23\pm0.10$\%.
This `signal,¹ however, is now gone; thus, the number in the table
as well as (\ref{eq:CLEOnc}) should be reduced by 0.0023 which, luckily,
is not a big change.

For $\Xi_c$, ALEPH and DELPHI used the CLEO measurement
of $Br(B\to \Xi_c)$ and added $Br(\Lambda_b\to\Xi_c)$
prediction by JETSET. However, the value used by ALEPH
$Br(B\to \Xi_c) = 3.9\pm1.5$\%~\cite{CLEOXic} is now superceded by 
$Br(B\to \Xi_c) = 2.0\pm1.0$\%~\cite{CLEOXic} 
which is used correctly by DELPHI.
The older value was based on the assumption that the semileptonic
rate of $\Xi_c$ is the same as that of $D$ which, together with
the measured $\Xi_c$ lifetime,
gave the branching
fraction of the hadronic mode used in the detection of $\Xi_c$.
However, the interference of the spectator $s$ quark and
the $c\to s$ transition was found to substantially 
enhance the semileptonic decay rate of $\Xi_c$~\cite{XicVolo}. 
The change in
$Br(B\to \Xi_c)$ largely reflects this correction. The CLEO value
of $n_c$ and that in the table 
as well as the DELPHI values already include this correction.
The ALEPH number, however, needs to be reduced; the corrected
number is
\beqa
  n_c &=& 1.211\pm0.036\pm0.035\pm0.053 \nonumber \\
    && \quad \hbox{(ALEPH, $\Xi_c$ corrected)}\,.
  \label{eq:ncALEPHcorr}
\eeqa
If we add the DELPHI values for $(c\bar c)$ and $\Xi_c$ to the
OPAL measurements for the rest of charmed hadrons, we obtain
\beqa
  n_c &=& 1.14\pm0.06\pm0.05 \nonumber \\
    && \hspace{-.3in}\hbox{(OPAL+$\Xi_c,(c\bar c)$ by DELPHI)}\,.
   \label{eq:ncOPALcorr}
\eeqa

The branching fractions of $D^0$, $D^+$, and $D_s^+$ used in the
charm counting are all normalized to $Br(D^0\to K^-\pi^+)$. Thus,
about 90\%\ of $n_c$ is controlled by it; namely, $n_c$ is
roughly inversely proportional to the value of $Br(D^0\to K^-\pi^+)$
used in the analyses. 
The 1996 particle data group
value is $Br(D^0\to K^-\pi^+) = 3.83\pm0.12$\%. Even though
the uncertainty is included in the systematic errors stated,
it is worthwhile to examine this number in some detail.
There is a new precise measurement by ALEPH~\cite{ALEPHKpi}
tagging $D^{*+}\to D^0\pi^+$ by the $p_T$ of the slow pion 
in jet, which gave
\[
  Br(D^0\to K^-\pi^+) = 3.897\pm0.094\pm0.117{\rm\%}\,.
\]
This is the method used by previous dominant measurements of
the quantity. A slightly different technique was employed by 
CLEO where the mode $B \to D^{*+}\ell\nu$ was used to 
tag $D^0$. This requires reconstruction of $B \to D^{*+}\ell\nu$
using the lepton and the slow pion from $D^{*+}$ only.
The figure~\ref{fg:CLEOKpi} shows the recoil mass distribution
of $\ell\pi$ pair for right-sign and wrong-sign samples.
There is a clear signal for the right-sign combinations.
\EPSFIGURE[h]{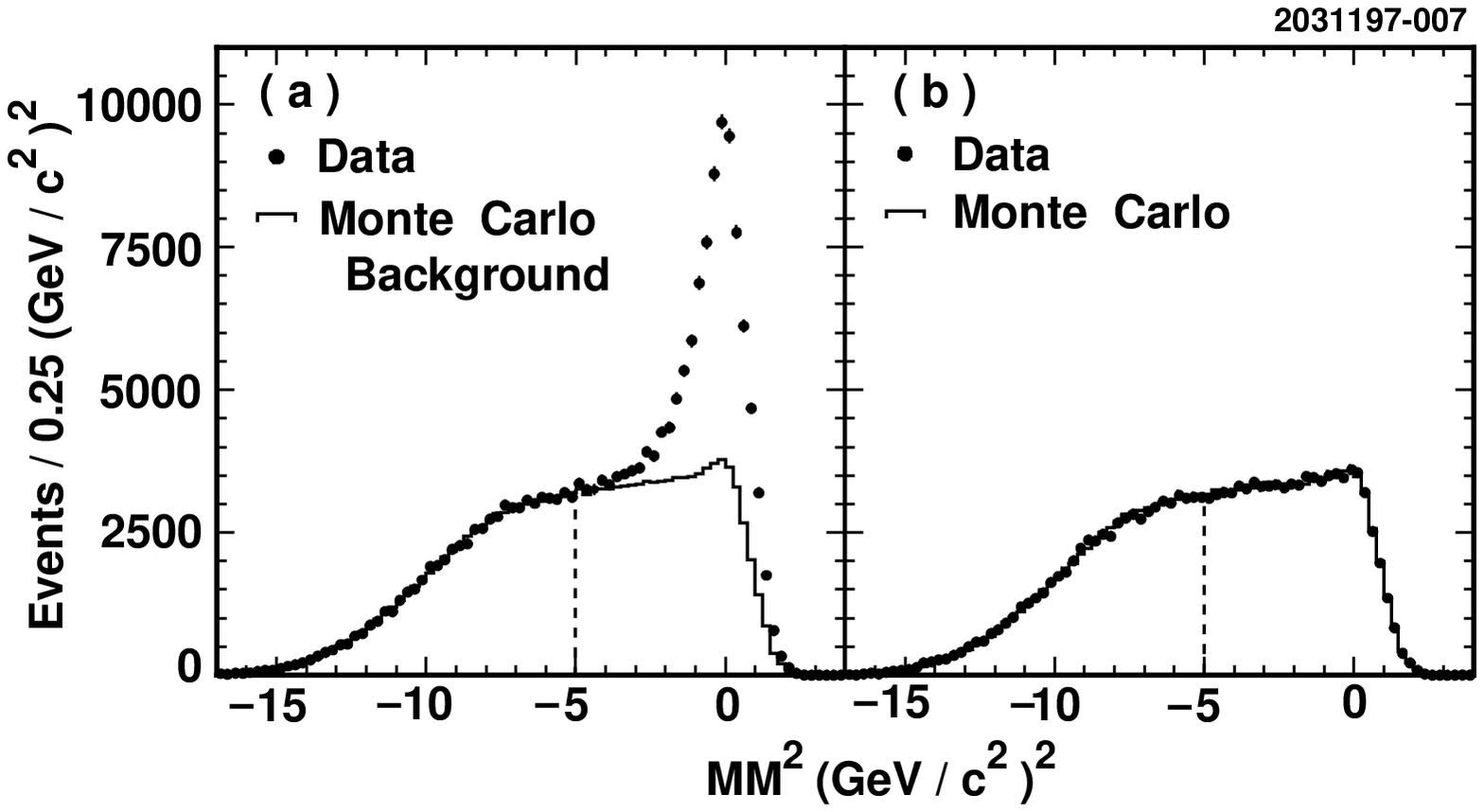,width=7cm}{The
      recoil mass distribution of $B \to D^{*+}\ell\nu$
      where the lepton and the soft pion only are detected.
      The right-sign [wrong-sign] $\ell\pi$ pairs are shown
      in (a) [(b)].
      \label{fg:CLEOKpi}}
The result was
\[
  Br(D^0\to K^-\pi^+) = 3.81\pm0.15\pm0.16{\rm\%}\,.
\]
These numbers are consistent with each other and there does not
seem to be a big problem in $Br(D^0\to K^-\pi^+)$.
There is also a CLEO measurement~\cite{CLEODl} 
of $Br(D^0\to K^-\pi^+)$ by
requiring that $B\to X\ell\nu$ is saturated by the
flavor-specific $D^{0,+}$ production apart from
$b\to u\ell\nu$ and $b\to D_s^+ X \ell\nu$ with the result
$Br(D^0\to K^-\pi^+) = 3.69\pm0.20$\%. Since this number is
obtained by forcing the charm counting in semileptonic sector
to come out correct, it is not suited to use in the charm
counting in general.

Another way to count the number of charm in $b$ decay is to
extract it from the vertex information. This has an advantage of
not dependent on the measured values of 
charm decay branching fractions.
One such analysis was performed by DELPHI~\cite{DELPHIncvtx} 
where the $b$ vertex tag was
used and the vertex information in the opposite side was used to
form the probability that all tracks with positive lifetimes come
from the primary vertex ($P^+_H$). Then, the distribution of
$-\log P^+_H$ was fit with monte-carlo shapes of $b\to 0c$,
$1c$, and $2c$ samples together with the $Z^0\to udsc$ backgrounds.
The $-\log P^+_H$ distribution for the 
1994 data and the result of the fit is
shown in the figure~\ref{fg:DELPHI02c}.
\EPSFIGURE[h]{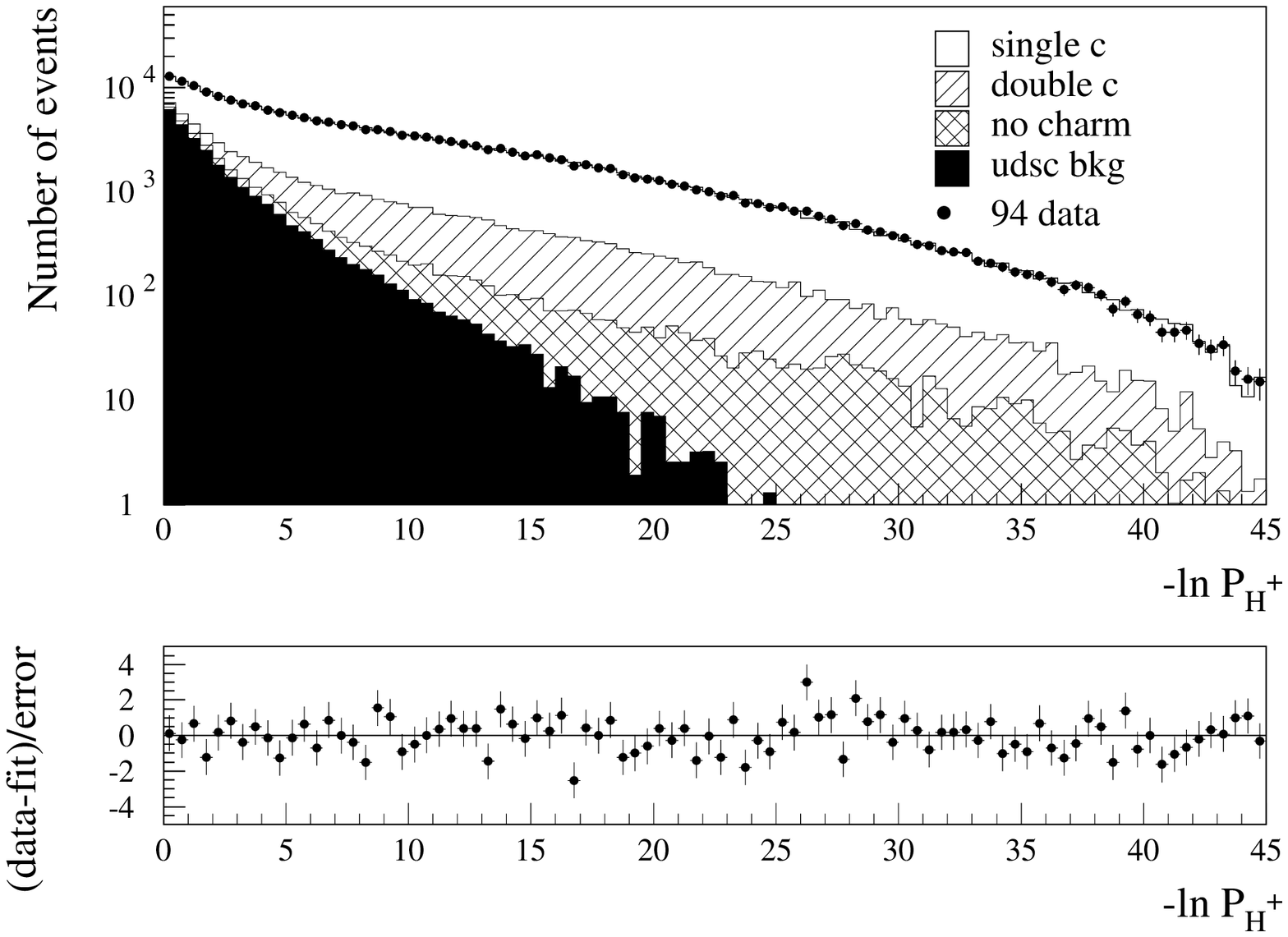,width=7cm}{Fit to the $-\log P^+_H$ 
       distribution (1994 data) by
       $b\to 0c$, $1c$, $2c$, and background, where $P^+_H$
       is the probability that all tracks with positive lifetimes come
       from the primary vertex. 
      \label{fg:DELPHI02c}}
The double charm and no-charm branching fraction thus obtained are
\beqa
     Br_{2c} &=& 0.136 \pm 0.042\,, \nonumber \\
     Br_{0c} &=& 0.033 \pm 0.021\,. \nonumber 
\eeqa
The zero-charm value includes the hidden-charm 
contribution $Br_{c\bar c}$ of
charmonia decays estimated to be $0.026\pm0.004$ which leaves
\[
   Br_{\rm rare} = 0.007\pm0.021
\]
as the `rare' branching fraction. 
This should be compared to 
the standard model prediction of $\rrare\Bsl = 0.026\pm0.011$.
The total number of charm was then estimated using $Br_{2c}$
only as
\beqa
   n_c &=& 1 + Br_{2c} + Br_{c\bar c} - \rrare\Bsl\nonumber \\
     &=& 1.147\pm0.041\pm0.008\,, \nonumber
  \label{eq:ncDELPHIvtx}
\eeqa
where $\rrare\Bsl = 0.016\pm0.08$ was used.

\section{Comparison of theory and experiment}

We will take (\ref{eq:LEPBsl}) as $\Bsl$ at $Z^0$ and convert it to
$\Upsilon(4S)$ value by multiplying $\tau_B/\tau_b$:
\[
    \Bsl = 11.07\pm 0.19\quad (Z^0, \hbox{corrected})\,. 
\]
For $\Bsl$ on $\Upsilon(4S)$, we use
the 1998 particle data
group value:
\beq
  \Bsl = 10.45\pm0.21 {\rm \%}\quad (\Upsilon(4S))\,.
\eeq
The discrepancy in $\Bsl$ between the $Z^0$ value and the
$\Upsilon(4S)$ value is then 2.2 $\sigma$.

For $n_c$ at $Z^0$, we will first take the average of the
ALEPH result corrected for $\Xi_c$ (\ref{eq:ncALEPHcorr}), the
OPAL result supplemented by DELPHI numbers for
$\Xi_c$ and charmonia (\ref{eq:ncOPALcorr}), and the DELPHI result
(\ref{eq:ncLEPexcl}) to obtain
\beq
   n_c = 1.178\pm0.035\pm0.054\quad(Z^0, {\rm exclusive})\,,
\eeq
where the last error is due to charm decay branching fractions.
Taking the average of this and the measurement using vertex
information (\ref{eq:ncDELPHIvtx}), we finally get
\beq
   n_c = 1.16\pm0.04\quad(Z^0)\,.
\eeq
The results are shown in the figure~\ref{fg:ncbsl} together with
\EPSFIGURE[h]{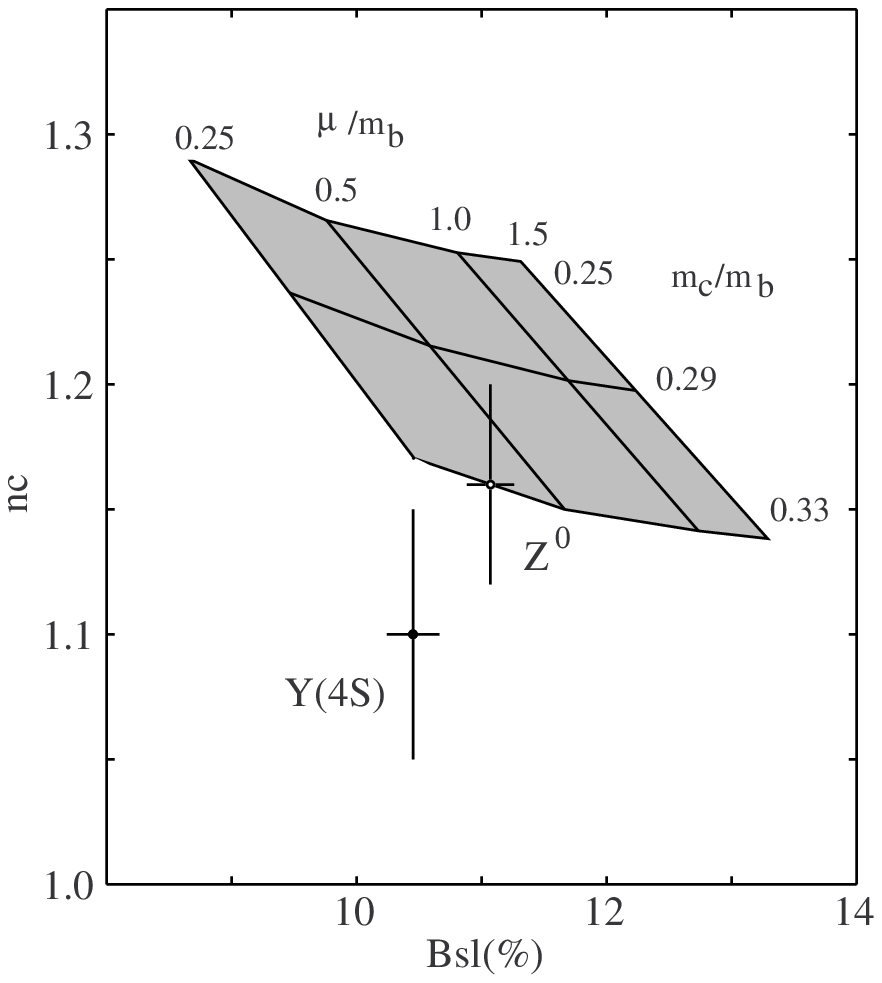,width=7cm}{Comparison of experiment and
      theory for $\Bsl$ and $n_c$. The measurement on $Z^0$ and
      those on $\Upsilon(4S)$ are shown separately. The value of
      $\Bsl$ on $Z^0$ is corrected to correspond to the average
      of $B^0$ and $B^+$.
      \label{fg:ncbsl}}
the theoretical expectation~\cite{NS}. Compared to a few years ago,
the $Z^0$ values have moved slightly toward the $\Upsilon(4S)$
values both in $\Bsl$ and $n_c$. The discrepancy 
between the measurements on $Z^0$ and those on $\Upsilon(4S)$ is still
uncomfortable. If one takes the measurement
on $\Upsilon(4S)$, the discrepancy between experiment and theory is
alarming, and such discrepancy would be eliminated if we assume
enhanced $\rrare$ beyond the value of the standard model which
would decrease the theoretical prediction of $\Bsl$ and also
decrease that of $n_c$.
If one takes the $Z^0$ values, however, the experiment and
theory are consistent.

\acknowledgments

\noindent The author would like to thank I. Dunietz for fuitful
collaboration and M. Neubert, E. Braun, and many
members of the CLEO collaboration for useful discussions.

\end {document}